\documentclass[9pt,conference]{IEEEtran}
\IEEEoverridecommandlockouts
\usepackage{cite}
\usepackage{amsmath,amssymb,amsfonts}
\usepackage{algorithmic}
\usepackage{graphicx}
\usepackage{textcomp}
\usepackage{xcolor}
\usepackage{hyperref}
\usepackage{bm}
\usepackage{amssymb}
\usepackage{booktabs}
\usepackage{multirow}
\usepackage[caption=false]{subfig} 
\def\BibTeX{{\rm B\kern-.05em{\sc i\kern-.025em b}\kern-.08em
    T\kern-.1667em\lower.7ex\hbox{E}\kern-.125emX}}
\begin{document}


\title{Incremental Disentanglement for Environment-Aware Zero-Shot Text-to-Speech Synthesis}

\author{\IEEEauthorblockN{Ye-Xin Lu, Hui-Peng Du, Zheng-Yan Sheng, Yang Ai$^*$\thanks{$^*$Corresponding author. This work was funded by the National Nature Science Foundation of China under Grant U23B2053 and 62301521, the Anhui Provincial Natural Science Foundation under Grant 2308085QF200, and the Fundamental Research Funds for the Central Universities under Grant WK2100000033.}, Zhen-Hua Ling} \IEEEauthorblockA{National Engineering Research Center of Speech and Language Information Processing, \\
University of Science and Technology of China, Hefei, P. R. China\\
{\small \tt \ \{yxlu0102, redmist, zysheng\}@mail.ustc.edu.cn, \{yangai, zhling\}@ustc.edu.cn}}}


\maketitle

\begin{abstract}
This paper proposes an Incremental Disentanglement-based Environment-Aware zero-shot text-to-speech (TTS) method, dubbed IDEA-TTS, that can synthesize speech for unseen speakers while preserving the acoustic characteristics of a given environment reference speech.
IDEA-TTS adopts VITS as the TTS backbone.
To effectively disentangle the environment, speaker, and text factors, we propose an incremental disentanglement process, where an environment estimator is designed to first decompose the environmental spectrogram into an environment mask and an enhanced spectrogram.
The environment mask is then processed by an environment encoder to extract environment embeddings, while the enhanced spectrogram facilitates the subsequent disentanglement of the speaker and text factors with the condition of the speaker embeddings, which are extracted from the environmental speech using a pretrained environment-robust speaker encoder.
Finally, both the speaker and environment embeddings are conditioned into the decoder for environment-aware speech generation.
Experimental results demonstrate that IDEA-TTS achieves superior performance in the environment-aware TTS task, excelling in speech quality, speaker similarity, and environmental similarity.
Additionally, IDEA-TTS is also capable of the acoustic environment conversion task and achieves state-of-the-art performance.
\end{abstract}


\vspace{-1mm}
\section{Introduction}
\label{sec: intro}
With the advancement of deep learning, text-to-speech (TTS) systems have rapidly evolved in recent years, enabling the synthesis of high-quality speech for both single and multiple speakers \cite{kim2020glow, kim2021conditional}.
More recently, there has been increased research interest in zero-shot TTS, which facilitates the synthesis of unseen speakers’ voices using only a few seconds of speaker reference speech \cite{casanova2021sc, casanova2022yourtts, wang2023neural}. 
These TTS models, trained on high-quality speech data, typically produce synthesized speech devoid of acoustic environmental characteristics.

In real-life scenarios, recorded speech often carries acoustic environmental characteristics influenced by factors such as recording room acoustics, quality and position of the recording devices, and ambient noise. 
Consequently, when speaker reference speech with environmental characteristics is used in zero-shot speech synthesis, the naturalness and speaker similarity of the synthesized speech would degrade.
To address this, noise-robust or environment-robust zero-shot TTS methods \cite{yang2023norespeech, fujita2024noise} have been developed to enable robust speech synthesis from distorted reference speech, typically by learning robust speaker embeddings \cite{fujita2024noise}. 
However, these methods do not disentangle the environmental factor from the reference speech, thus limiting their capability to only generate high-quality raw speech.

In certain applications such as audiobooks, virtual meetings, and virtual reality, there is a need to synthesize speech with specific environmental characteristics.
Previous studies on acoustic environment conversion \cite{su2020acoustic, im2024diffrent} utilized environment encoders to extract environment embeddings, enabling the transformation of speech from one environment to another. 
However, these methods can only alter the acoustic environment without modifying the speech content, which restricts their applicability.
Environment-aware TTS aimed to generate speech suited to specific acoustic environments, but it still faced several challenges.
The previous approach \cite{tan2021environment} attempted to disentangle environmental and speaker factors simultaneously, leading to entanglement between the two, which posed difficulties when dealing with unseen environments and speakers. 
Additionally, it solely focused on room reverberation, making it less effective for application in diverse real-world acoustic environments.

Therefore, in this paper, we propose IDEA-TTS, which incorporates incremental disentanglement for environment-aware zero-shot TTS, enabling the generation of speech for unseen speakers with controllable acoustic environments.
Built upon the VITS framework, IDEA-TTS first disentangles environmental factors, followed by speaker factors, considering the global influence of environmental characteristics on speaker features.
Specifically, an environment estimator is first introduced to decompose the environmental spectrogram into an environment mask and an enhanced spectrogram. 
The environment mask is used to extract the environment embeddings, while the enhanced spectrogram is further used to disentangle speaker and text factors, conditioned on speaker embeddings extracted from the environmental speech via a pretrained environment-robust speaker encoder.
Finally, both the speaker and environment embeddings are conditioned into the decoder to generate speech with the corresponding acoustic environment and speaker characteristics.
Both subjective and objective results indicate that IDEA-TTS effectively disentangles environmental, speaker, and text factors, achieving promising results in environment-aware TTS synthesis and state-of-the-art (SOTA) performance in acoustic environment conversion.

\vspace{-1mm}
\section{Methodology}
\label{sec: method}
\begin{figure*}[ht]
\centering
    \begin{minipage}[b]{0.49\textwidth}
        \centering
        \subfloat[Training procedure]{\includegraphics[height=7.5cm]{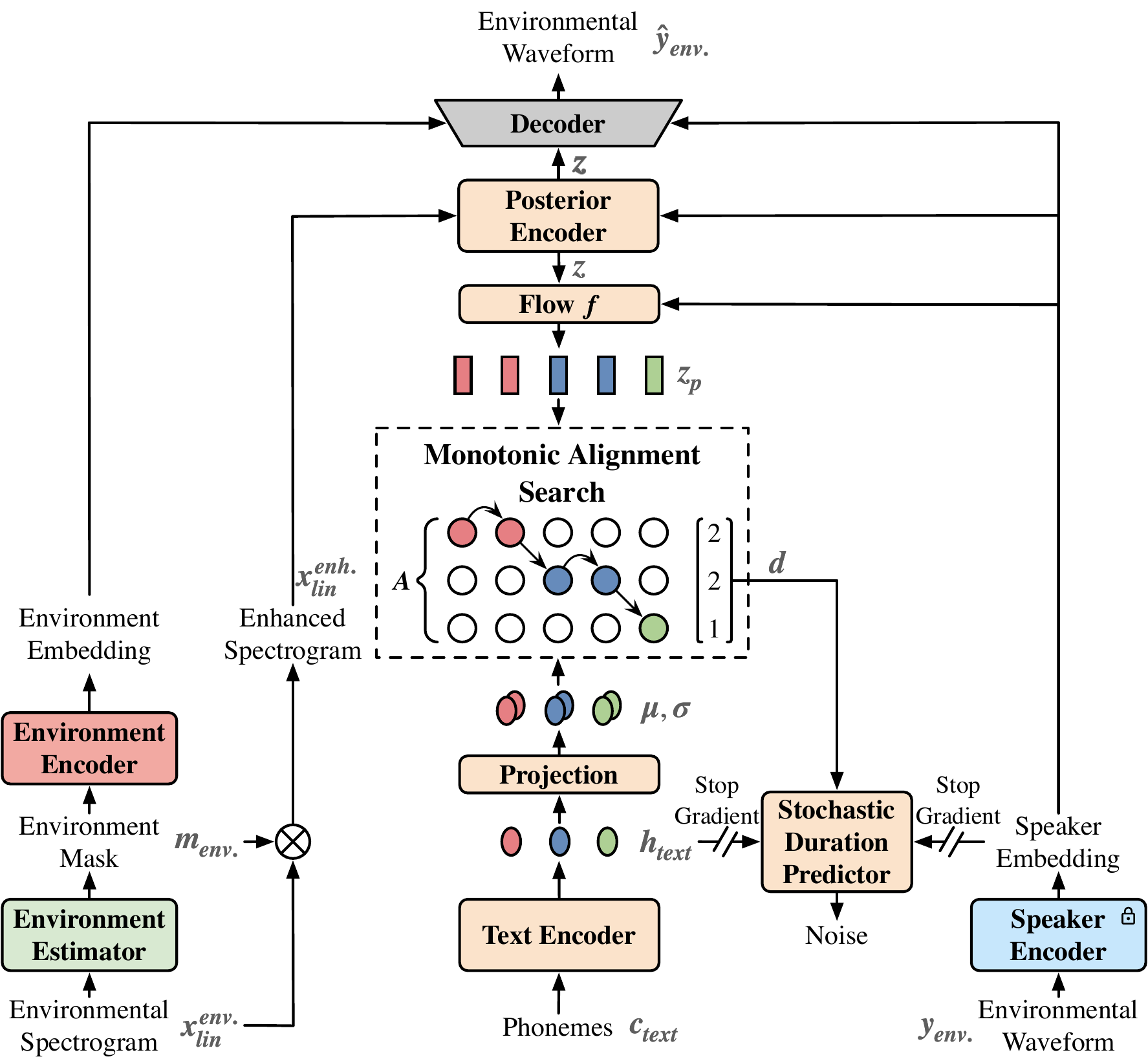}\label{fig: train}}
    \end{minipage}
    \hfill
    \begin{minipage}[b]{0.49\textwidth}
        \centering
        \subfloat[Inference procedure]{\includegraphics[height=7.5cm]{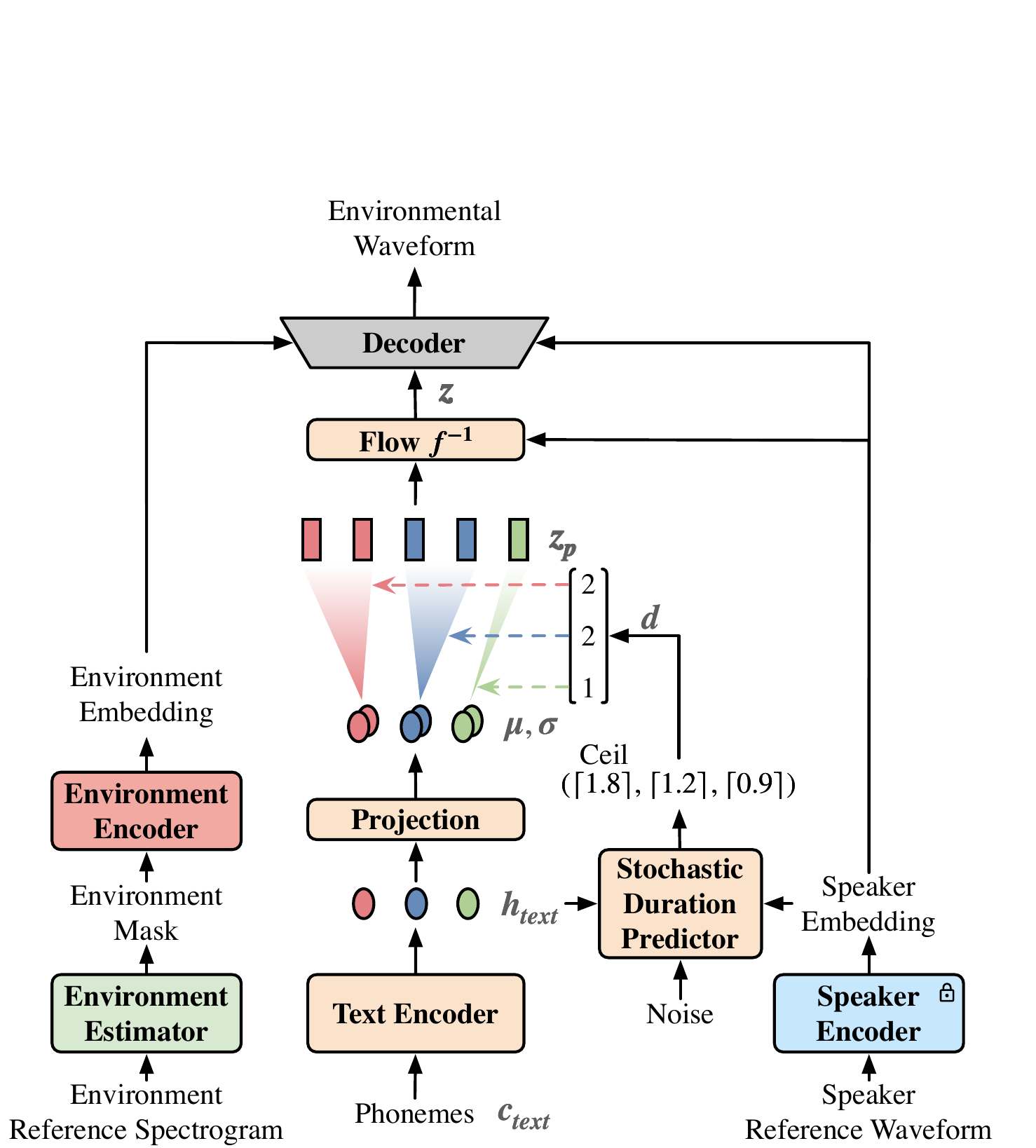}\label{fig: infer_tts}}
    \end{minipage}
    \caption{Overview of the model structure, training procedure, and inference procedure of the proposed IDEA-TTS.}
    \label{fig: model}
\end{figure*}

The overall structure of the proposed IDEA-TTS is illustrated in Fig.~\ref{fig: model}.
The backbone of IDEA-TTS is inherited from VITS \cite{kim2021conditional}, which proposes a conditional variational autoencoder (CVAE) with adversarial learning for end-to-end TTS.
VITS comprises a posterior encoder and a decoder for spectrogram encoding and waveform reconstruction, a text encoder for processing input phonemes, a normalizing flow for improving the flexibility of the prior distribution, and a stochastic duration predictor (SDP) for predicting the phoneme duration.

Based upon the VITS model, our proposed IDEA-TTS introduces an environment estimator to decompose the environmental spectrogram into an environment mask and an enhanced spectrogram.
Additionally, it incorporates a speaker encoder and an environment encoder to respectively extract speaker and environment embeddings, which are ultimately conditioned to the decoder to generate environment-aware speech. 
The details of the environment estimator, environment encoder, and speaker encoder, along with the training and inference procedures are described as follows.

\subsection{Model Structure}
\subsubsection{Environment Estimator}
The environment estimator aims to assess the discrepancy between environmental speech and raw speech, thereby facilitating the effective disentanglement of the environmental factor.
Inspired by the spectral masking method \cite{srinivasan2006binary, narayanan2013ideal} commonly used in speech enhancement, we design the environment estimator to predict an environment mask $\bm{m}_{env.}$ from the environmental spectrogram $\bm{x}_{lin}^{env.}$ and multiply it with $\bm{x}_{lin}^{env.}$ to obtain an enhanced spectrogram $\bm{x}_{lin}^{enh.}$.
The environment mask, considered to contain all the environmental information, is used as input to the environment encoder to extract the environment embedding, while the environment-independent enhanced spectrogram is fed into the posterior encoder to further disentangle speaker and text factors.

Given the recent widespread use of multi-head attention \cite{vaswani2017attention} in speech enhancement for capturing long-term dependencies, we design a transformer \cite{vaswani2017attention} based environment estimator, which comprises an input convolutional layer, multiple transformer layers, an output convolutional layer, and a parametric rectified linear unit (PReLU) activation \cite{he2015delving} function to predict the environment mask.
Additionally, we employ power-law compression with a compression factor of 0.3 to enhance the prediction of the environment mask by reducing the dynamic range of the spectrogram as in our previous work \cite{lu2023mp}.

\subsubsection{Environment Encoder}
The environment encoder is designed to extract an environment embedding that exclusively captures environmental information. 
Therefore, we use the environment mask as the input to the environment encoder, rather than the entire environmental spectrogram as employed in previous works \cite{tan2021environment, im2024diffrent}. 
We posit that the environment mask contains minimal speech content and more effectively reflects the acoustic environment.
Given the assumption that acoustic environmental characteristics are time-invariant, we employ the ECAPA-TDNN \cite{desplanques2020ecapa}, a speaker embedding extraction network with strong static feature extraction capabilities, as the environment encoder to extract fixed-length environment embeddings from variable-length environment masks.

\subsubsection{Speaker Encoder}
To extract speaker embeddings from environmental speech, we follow \cite{tan2021environment} to use a pretrained environment-robust speaker encoder.
Specifically, we adopt the H/ASP model \cite{heo2020clova} as the speaker encoder as in \cite{casanova2022yourtts}. 
The H/ASP model is trained on the VoxCeleb2 dataset \cite{chung2018voxceleb2}, which includes a large amount of speech data from over 6,000 speakers, recorded across various acoustic environments. 
Consequently, this pretrained speaker encoder is well-suited to effectively model speaker-dependent factors while minimizing the influence of environment-dependent factors.

\subsection{Training Procedure}
In the training stage, as illustrated in Fig.~\ref{fig: train}, the posterior encoder uses the enhanced linear spectrogram $\bm{x}_{lin}^{enh.}$ as input to encode an environment-independent latent variable $\bm{z}$.
On the one hand, the latent variable $\bm{z}$ is fed into the decoder, together with the environment embedding and speaker embedding to reconstruct the environmental speech waveform.
On the other hand, it is transformed into $\bm{z}_p$ through a normalizing flow $f$ with the condition of the speaker embedding, and then aligned with the output of the text encoder using monotonic alignment search (MAS) \cite{kim2020glow}.
Additionally, the SDP estimates the distribution of the phoneme duration from the output of the text encoder, also conditioned on the speaker embedding.

For training IDEA-TTS, we follow the adversarial training criteria of VITS \cite{kim2021conditional}, jointly training the environment estimator 
with the TTS module using an additional spectrogram enhancement loss:
\begin{equation}
	\mathcal{L}_{SE} = \big\Vert \bm{x}_{lin}^{raw} - \bm{x}_{lin}^{enh.} \big\Vert_{\mathrm{F}}^2 + \big\Vert \bm{x}_{mel}^{raw} - \bm{x}_{mel}^{enh.} \big\Vert_{1},
\end{equation}
where $\Vert\cdot\Vert_{\mathrm{F}}$ and $\Vert\cdot\Vert_{1}$ denote the Frobenius norm and the L1 norm, respectively; $\bm{x}_{lin}^{raw}$ and $\bm{x}_{lin}^{enh.}$ denote the raw linear spectrogram and the enhanced linear spectrogram, respectively; and $\bm{x}_{mel}^{raw}$ and $\bm{x}_{mel}^{enh.}$ denote their corresponding mel-spectrograms.

Since the latent variable $\bm{z}$ is environment-independent, the decoder is required to generate environment-aware speech when the environment embedding is conditioned, and raw speech when it is not (given the all-zero environment embedding). 
Both generated environmental and raw speech are used for loss computation and fed to the discriminator. 
This further supervises the disentanglement of the environmental factor, enabling the model to achieve environment-robust TTS without requiring additional raw reference speech.

\subsection{Inference Procedure}
The inference procedure for the environment-aware TTS synthesis is illustrated in Fig.~\ref{fig: infer_tts}.
The environment reference spectrogram is processed by the environment estimator and environment encoder to obtain the environment embedding, while the speaker embedding is extracted from a speaker reference waveform. 
The text encoder predicts the distribution $P_{\bm{z}_p}$ from phonemes, and $\bm{z}_p$ is sampled from $P_{\bm{z}_p}$ following the duration $\bm{d}$, which is sampled from random noise through the inverse transformation of SDP.
Finally, $\bm{z}_p$ is transformed into $\bm{z}$ through the inverse normalizing flow $f^{-1}$ and then combined with the environment embedding and speaker embedding in the decoder to generate the environment-aware speech waveform.

\section{Experiments}
\label{sec: exp}
\subsection{Dataset and Experimental Setup}
We conducted experiments on the DDS dataset \cite{li2021dds}, which comprised 12 hours of raw speech data from 48 speakers (24 females and 24 males) and corresponding environmental speech data from 27 different recording conditions. 
Each condition included recordings captured from 6 microphone positions to simulate varying levels of noise and reverberation.
In total, the DDS dataset provides approximately 2,000 hours (9 rooms $\times$ 3 devices $\times$ 6 positions $\times$ 12 hours) of realistic environmental speech data.
For the test set, we followed \cite{im2024diffrent} to select 4 unseen speakers (2 females and 2 males) and 1 unseen recording environment (livingroom1, Uber microphone, F position).

All audio samples were resampled to 16 kHz. 
We followed the model configuration of VITS \cite{kim2021conditional}. 
The linear spectrogram and mel-spectrogram were computed using short-time Fourier transform (STFT), with the FFT point number, window size, and hop size set to 1024, 1024, and 256, respectively. 
All the models were trained for 800k steps on an NVIDIA A100 GPU with a batch size of 64, using the AdamW optimizer \cite{loshchilov2017decoupled} with $\beta_1=0.8$, $\beta_2=0.99$, and a weight decay of $\lambda=0.01$.
The learning rate was initially set to $2.0 \times 10^{-4}$ and scheduled to decay by a factor of $0.999^{1/8}$ every epoch.
\footnote{Audio samples of the proposed IDEA-TTS can be accessed at \href{https://yxlu-0102.github.io/IDEA-TTS}{https://yxlu-0102.github.io/IDEA-TTS}.}

\subsection{Baselines and Evaluation Metrics}
\subsubsection{Text-to-Speech Tasks}
We evaluated the IDEA-TTS on two TTS tasks: environment-robust TTS and environment-aware TTS.
The environment-robust TTS task is used to evaluate the models' capabilities to extract robust speaker embedding from environmental speaker reference speech and synthesize high-quality output.
For this task, the comparison involved a baseline YourTTS \cite{casanova2022yourtts}, which is a zero-shot TTS model trained on the raw speech data of the DDS dataset, against the proposed IDEA-TTS and its variant IDEA-TTS (w/o ID), which performed simultaneous disentanglement as the previous environment-aware TTS approach \cite{tan2021environment}.
IDEA-TTS (w/o ID) omitted the environment estimator, and environment embedding was extracted directly from the environmental spectrogram and conditioned alongside the speaker embedding into the posterior encoder, normalizing flow, and decoder.
For IDEA-TTS (w/o ID) in environment-robust TTS, the raw speech was used as the environment reference.
For the environment-aware TTS task, since YourTTS was only capable of raw speech generation, we used the two proposed methods for comparison.

For the evaluation of the TTS task, we selected the test set with a minimum text length of 30 characters.
Using unseen speakers, we chose one speech sample from an unseen environment as the speaker reference and another random sample from a seen environment as the environment reference. 
Both samples had a duration of at least 3 seconds.
We conducted both objective and subjective evaluations.
For objective evaluation, we used speaker encoder cosine similarity (SECS) to evaluate speaker similarity and character error rate (CER) to assess speech intelligibility.
SECS was calculated using a SOTA speaker verification model, WavLM-TDNN \cite{chen2022wavlm} \footnote{\href{https://huggingface.co/microsoft/wavlm-base-plus-sv}{https://huggingface.co/microsoft/wavlm-base-plus-sv}.}, to extract speaker vectors.
Its values range from -1 to 1, where higher values indicate greater similarity. 
CER was calculated using an automatic speech recognition (ASR) model, FunASR \cite{gao2023funasr} \footnote{\href{https://github.com/modelscope/FunASR}{https://github.com/modelscope/FunASR}.}, to transcribe the generated speech.
For subjective evaluation, we crowd-sourced over 30 raters on Amazon Mechanical Turk to evaluate the naturalness, speaker similarity, and environment similarity of 20 sets of speech samples, using 5-scale mean opinion score (MOS), speaker similarity MOS (SSMOS), and environment similarity MOS (ESMOS), respectively.
The MOS-based results were reported with 95\% confidence intervals.

\subsubsection{Acoustic Environment Conversion Task}
For IDEA-TTS, the incremental disentanglement process ensures that the latent variable $\bm{z}$ is environment-independent, enabling it to perform the conversion of acoustic environments through the combined use of selected modules.
Specifically, the source environmental spectrogram undergoes the environment estimator and posterior encoder to obtain a latent variable $\bm{z}$ that only contains speech content.
The target environmental spectrogram is processed through the environment estimator and environment encoder to obtain the target environment embedding. 
Finally, the latent variable $\bm{z}$ together with the target environment embedding and the source speaker embedding, is fed into the decoder to produce the speech waveform with the target environment while containing the speech content of the source speech.
However, for IDEA-TTS (w/o ID) with simultaneous disentanglement, the distribution predicted by the text encoder is environment-independent, allowing for acoustic environment conversion through an additional inverse and forward process of the normalizing flow.

For evaluation, we compared the two proposed models with the ``W-R2-C'' model of the DiffRENT \cite{im2024diffrent} in three cases: (1) ``Env-to-Clean'': convert the unseen environment to the clean environment; (2) ``Clean-to-Env'': convert the clean environment to the unseen environment; (3) ``Env-to-Env'': convert the unseen environment to a randomly chosen seen environment.
For each case, we randomly chose 500 pairs of source and target environmental speech as in \cite{im2024diffrent}. 
For the evaluation of the acoustic environment conversion task, we adopted the log-spectral distance (LSD) \cite{gray1976distance}, perceptual evaluation of speech quality (PESQ) \cite{rix2001perceptual}, and the virtual speech quality objective listener (ViSQOL) \cite{chinen2020visqol} as the objective metrics.
ViSQOL used a spectral-temporal measure of similarity between a reference and a synthesized speech to produce a 5-scale objective MOS score.

\section{Results and Analysis}
\label{sec: result}
\subsection{Results on the Text-to-Speech Tasks}
The experimental results of the environment-robust TTS are shown in Table~\ref{tab: environment-robust}. 
Compared to YourTTS with raw speaker reference speech (denoted as YourTTS (Raw Ref.)), YourTTS with environmental speaker reference showed a slight decline in speech naturalness and speaker similarity, as indicated by MOS, SSMOS, and SEC, but a marginal improvement in speech intelligibility, as indicated by CER, validating the environmental robustness of the pretrained speaker encoder.
IDEA-TTS performed comparably to YourTTS (Raw Ref.) in both subjective and objective evaluations, demonstrating effective disentanglement of speaker factors and high-quality environment-robust speech synthesis. 
When comparing different disentanglement approaches, IDEA-TTS (w/o ID) significantly underperformed IDEA-TTS across all metrics.
Since they utilized the same speaker embeddings, this suggested that under simultaneous disentanglement, the environmental information interfered with the disentanglement of the speaker factor, resulting in a lower speaker similarity in the synthesized speech.
Furthermore, the insufficient disentanglement of environmental and speaker factors ultimately affected the disentanglement of text factors, leading to poor performance of IDEA-TTS (w/o ID) in terms of speech naturalness and intelligibility.

The experimental results of the environment-aware TTS are depicted in Table~\ref{tab: environment-aware}.
Overall, the results of the two proposed models in terms of speech naturalness, intelligibility, and speaker similarity were consistent with those observed in the environment-robust TTS task.
Considering the environmental similarity, IDEA-TTS still excelled over IDEA-TTS (w/o ID) in terms of ESMOS, which further suggested that, under simultaneous disentanglement, environmental and speaker information tended to be entangled, hindering the disentanglement of each factor and ultimately limiting the environment-aware speech generation capability of IDEA-TTS (w/o ID).
Therefore, incremental disentanglement is a more reasonable and effective solution for environment-aware TTS.

\begin{table}[t!]\footnotesize
  \caption{Comparison of Subjective and Objective Evaluation Results for the Environment-Robust TTS Task}
  \label{tab: environment-robust}
  \centering
  \resizebox{\linewidth}{!}{
  \begin{tabular}{l|cccc}
    \toprule
    Method & MOS & SSMOS & SECS & CER (\%) \\
    \midrule
    Ground Truth        & 3.95$\pm$0.06 & - & 1.000  & 1.83 \\
    \midrule
    YourTTS (Raw Ref.)  & \textbf{3.90$\pm$0.06} & 3.82$\pm$0.07 & \textbf{0.873} & 5.95 \\
    YourTTS \cite{casanova2022yourtts}           & 3.84$\pm$0.06          & 3.77$\pm$0.07 & 0.867 & 5.68 \\
    \midrule
    IDEA-TTS (w/o ID)   & 3.76$\pm$0.07          & 3.56$\pm$0.08 & 0.847 & 12.23 \\
    IDEA-TTS            & \textbf{3.90$\pm$0.06} & 3.84$\pm$0.07 & \textbf{0.873} & \textbf{5.26} \\
    \bottomrule
  \end{tabular}}
\end{table}
\begin{table}[t!]\Huge
  \caption{Comparison of Subjective and Objective Evaluation Results for the Environment-Aware TTS Task}
  \label{tab: environment-aware}
  \centering
  \resizebox{\linewidth}{!}{
  \begin{tabular}{l|ccccc}
    \toprule
    Method & MOS & SSMOS & ESMOS & SECS & CER (\%) \\
    \midrule
    Ground Truth        & 3.65$\pm$0.08 & - & - & 0.925 & 2.87 \\
    \midrule
    IDEA-TTS (w/o ID)   & 3.45$\pm$0.09 & 3.65$\pm$0.07 & 3.67$\pm$0.07 & 0.830 & 12.80  \\
    IDEA-TTS            & \textbf{3.65$\pm$0.07} & \textbf{3.78$\pm$0.07} & \textbf{3.75$\pm$0.08} & \textbf{0.845} & \textbf{5.93} \\ 
    \bottomrule
  \end{tabular}}
\end{table}

\subsection{Results on the Acoustic Environment Conversion Task}
The experimental results of the acoustic environment conversion task are depicted in Table~\ref{tab: AEC}.
Overall, the proposed IDEA-TTS achieved a SOTA performance, demonstrating its strong capability for environmental disentanglement.
Compared to the baseline method DiffRENT, in the ``Env-to-Clean'' scenario, IDEA-TTS significantly outperformed DiffRENT especially in PESQ, indicating its superior speech enhancement capabilities. 
In the other two scenarios, IDEA-TTS was on par with DiffRENT in terms of LSD but surpassed it in ViSQOL, indicating that the speech converted by IDEA-TTS had a higher spectral-temporal similarity with the target speech.
Conversely, IDEA-TTS (w/o ID) experienced a performance collapse across all the metrics, once again verifying that simultaneous disentanglement resulted in information entanglement, thereby affecting the quality of the converted speech.

\begin{table}[t!]\Huge
  \caption{Comparison of Objective Evaluation Results for the Acoustic Environment Conversion Task}
  \label{tab: AEC}
  \centering
  \resizebox{\linewidth}{!}{
  \begin{tabular}{l|ccc|cc|cc}
    \toprule
    \multirow{2}{*}{Method} & \multicolumn{3}{c|}{Env-to-Clean} & \multicolumn{2}{c|}{Clean-to-Env} & \multicolumn{2}{c}{Env-to-Env} \\
    \cmidrule{2-8}
                      & LSD  & PESQ & ViSQOL & LSD  & ViSQOL & LSD  & ViSQOL \\
    \midrule
    Unprocessed       & 1.24 & 1.34 & 3.03   & 1.24 & 3.06   & 0.95 & 3.24  \\
    \midrule
    DiffRENT \cite{im2024diffrent}  & 0.92 & 1.53 & 3.56   & \textbf{0.94} & 3.26   & \textbf{0.91} & 3.53  \\
    \midrule
    IDEA-TTS (w/o ID) & 1.15 & 1.22 & 2.89   & 1.02   & 2.72 & 1.01 & 2.97  \\
    IDEA-TTS          & \textbf{0.89} & \textbf{1.86} & \textbf{3.68} & \textbf{0.94} & \textbf{3.33} & \textbf{0.91} & \textbf{3.55} \\
    \bottomrule
  \end{tabular}}
\end{table}
\begin{figure}[t!]
\centering
    \begin{minipage}[b]{0.49\linewidth}
        \centering
        \subfloat[Simultaneous disentanglement]{\includegraphics[width=\linewidth]{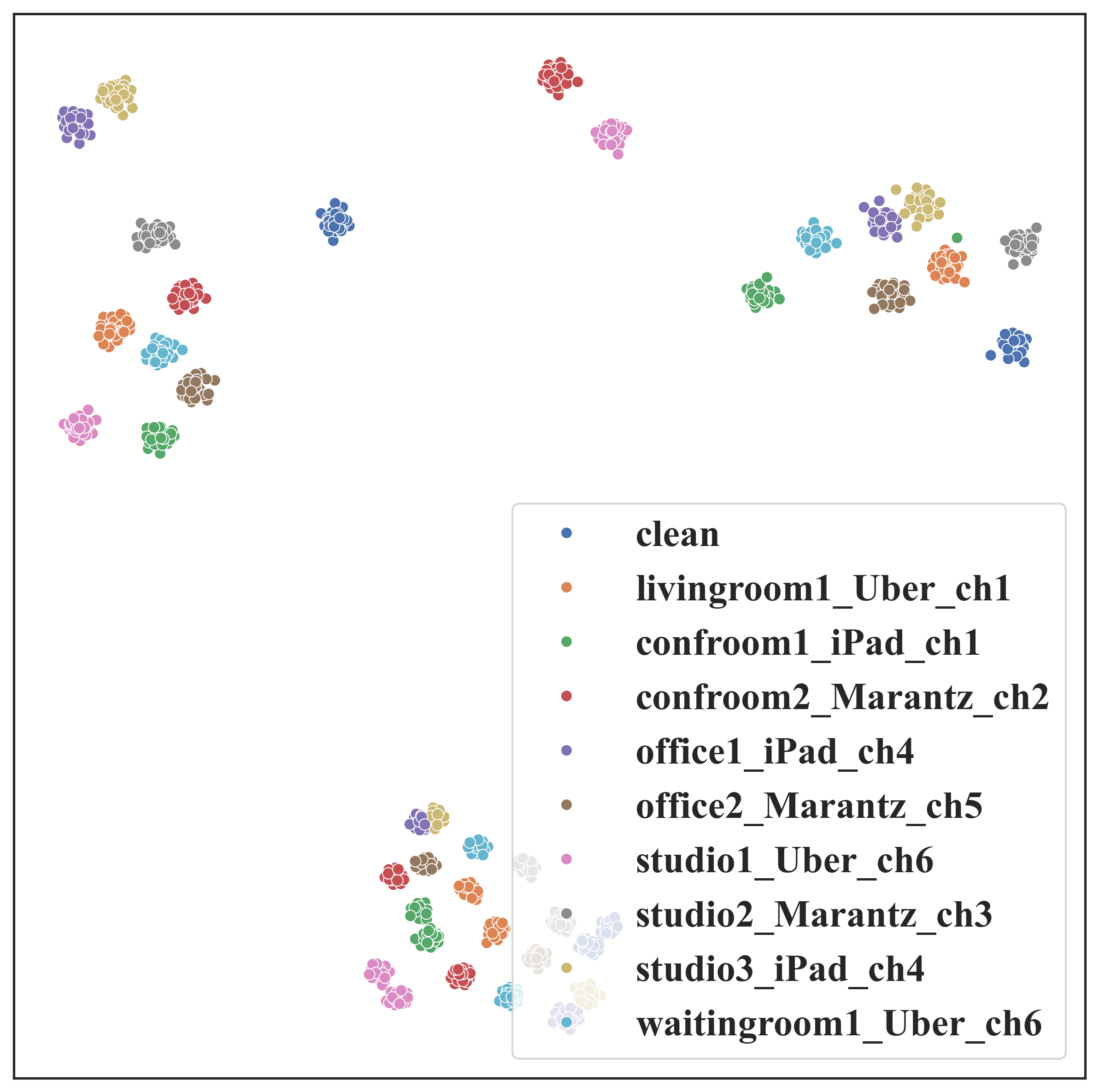}\label{fig: t-sne-sd}}
    \end{minipage}
    \hfill
    \begin{minipage}[b]{0.49\linewidth}
        \centering
        \subfloat[Incremental disentanglement]{\includegraphics[width=\linewidth]{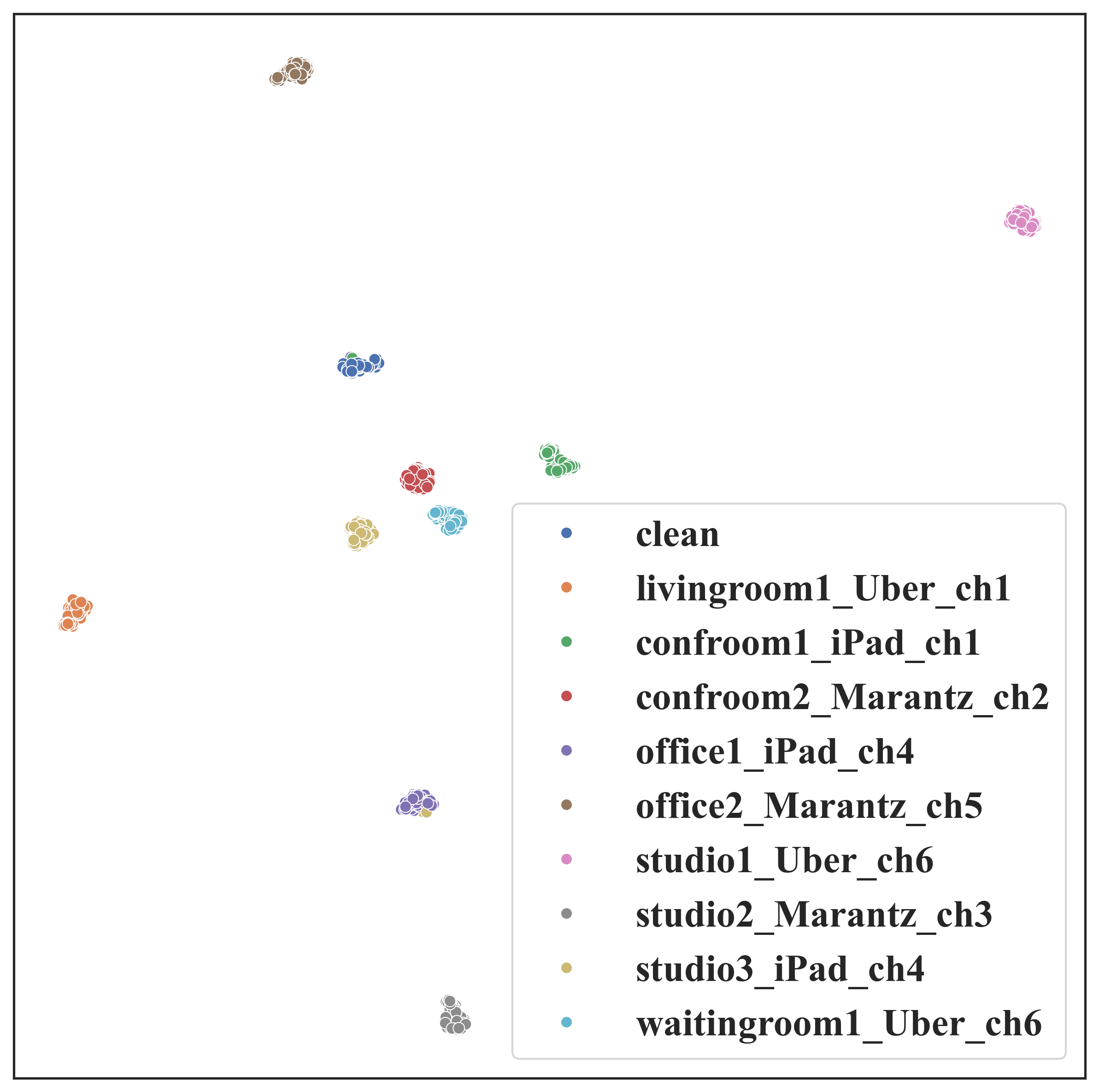}\label{fig: t-sne-id}}
    \end{minipage}
    \caption{t-SNE scatter visualization of the environment embeddings under different disentanglement approaches, where different colors correspond to different acoustic environments, and each point represents a speech sample.}
    \label{fig: t-sne}
\end{figure}

\subsection{Analysis of the Environment Encoder}
To compare the performance of the environment encoders trained with different disentanglement methods, we visualized the environment embeddings extracted by these encoders using t-SNE. 
We used speech samples from 10 different acoustic environments (including one clean environment, one unseen environment, and eight seen environments), with 100 speech samples per environment, drawn from four unseen speakers.
Fig.~\ref{fig: t-sne-sd} and Fig.~\ref{fig: t-sne-id} showed the results of the simultaneous disentanglement from IDEA-TTS (w/o ID) and the incremental disentanglement from IDEA-TTS, respectively.
It is clear that the environment embeddings obtained through incremental disentanglement form more distinct clusters, while those from simultaneous disentanglement tend to overlap and mix together.

\section{Conclusions}
In this paper, we proposed IDEA-TTS, an environment-aware zero-shot TTS model that incrementally disentangled acoustic environment and speaker factors. 
Built upon the VITS framework, the core of IDEA-TTS lay in the incremental disentanglement process, where an environment estimator was designed to estimate an environment mask from the environment spectrogram, thereby disentangling the acoustic environment from the speech content. 
The enhanced spectrogram, devoid of environmental information, was further used to disentangle the speaker and text factors conditioned on the speaker embeddings extracted by a pretrained environment-robust speaker encoder. 
As a result, IDEA-TTS effectively disentangled environment, speaker, and text factors, allowing for the generation of speech for unseen speakers with controllable acoustic environments. 
Both subjective and objective experimental results demonstrated the superiority of the proposed method.
In future work, we will incorporate acoustic environment control into large-language-model-based TTS.

\bibliographystyle{IEEEtran}
\bibliography{refs}

\end{document}